\pdfoutput=1
\documentclass[11pt,english]{article}
\usepackage[T1]{fontenc}
\usepackage[latin9]{inputenc}
\usepackage{nicefrac}
\usepackage{graphicx}
\usepackage{setspace}
\usepackage{amssymb}

\usepackage{subfigure}

\usepackage{nicefrac}

\usepackage{esint}

\usepackage{hyperref}

\usepackage{babel}

\begin{document}

\title{Algorithmic complexity and randomness in elastic solids}

\author{J. Ratsaby$^{\dagger}$ and J. Chaskalovic$^{\ddagger}$}

\maketitle
$^{\dagger}$Department of Electrical and Electronics Engineering,
$^{\ddagger}$Department of Mathematics and Computer Science, Ariel
University Center, Ariel 40700, ISRAEL and IJLRDA, University Pierre
and Marie Curie - Paris VI, FRANCE

$^{\dagger}$ratsaby@ariel.ac.il, $^{\ddagger}$jch@ariel.ac.il 

\begin{abstract}
A system comprised of an elastic solid and its response to an external
random force sequence is shown to behave based on the principles of
the theory of algorithmic complexity and randomness. The solid distorts
the randomness of an input force sequence in a way proportional to
its algorithmic complexity. We demonstrate this by numerical analysis
of a one-dimensional vibrating elastic solid (the system) on which
we apply a maximally random input force. The level of complexity of
the system is controlled via external parameters. The output response
is the field of displacements observed at several positions on the
body. The algorithmic complexity and stochasticity of the resulting
output displacement sequence is measured and compared against the
complexity of the system. The results show that the higher the system
complexity the more random-deficient the output sequence. This agrees
with the theory introduced in \cite{Ratsaby_entropy} which states
that physical systems such as this behave as algorithmic selection-rules
which act on random actions in their surroundings.

\end{abstract}

\section{Introduction}

Consider an elastic beam having a length $L$, (for instance, a bridge).
It has some finite descriptive complexity consisting of all the information
contained in the engineering design documents. These documents can
be put into a single computer file that can be represented by a finite
binary string $z$. This binary sequence has an algorithmic complexity
which is defined as the length of the shortest computer program that
can generate the sequence. This is defined as the Kolmogorov complexity
$K(z)$ of the string $z$ (see \cite{Kolmogorov65}). Now consider
a \emph{random} input force sequence applied at one of the two ends
of the bridge, for instance, suppose there is a person jumping up
and down sporadically on the bridge at its entrance (position $0$).
Denote by $x$ the binary sequence representing this up/down symbols
over some fixed time-interval $[0,T]$. Intuitively, being that $x$
is random makes its complexity $K(x)$ maximal and hence close to
its actual length $\ell(x)$ since there is no redundancy in the patterns
of $x$ that can be used to compress it significantly below its length.
Now consider an observer which measures the displacements on the beam
at its other end (position $L$). He records this over the time interval
$[0,T]$ and compares it to a fixed threshold thereby producing a
binary output sequence $y$ consisting of up/down symbols that represent
the movement of the beam at position $L$. This sequence has a finite
algorithmic complexity $K(y)$. In this paper we show that for such
a physical system, the system complexity $K(z)$, the output complexity
$K(y)$ and its level of randomness are all related and there exist
statistically significant correlations between them.

Ratsaby \cite{Ratsaby_entropy} introduced a quantitative definition
of the information content of a static structure (a solid) and explained
its relationship to the stability and symmetry of the solid. His model
is based on concepts of the theory of algorithmic information and
randomness. He modeled a solid as a selection rule of a finite algorithmic
complexity which acts on an incoming random sequence of particles
in the surroundings. This selection mechanism is intrinsically connected
to the solid's complex non-linear structure (partly a consequence
of its internal atomic vibrations) and its intricate time-response
to external stimulus. As postulated in \cite{Ratsaby_entropy}, a
simple solid is one whose information content is small. Its selection
behavior is of low complexity since it can be described by a more
concise time-response model (shorter computer program). The solid's
stability over time is explained in \cite{Ratsaby_entropy} by using
the stochastic property of the frequency stability of a random sequence.
Accordingly, the physical stability of the system (solid) is intrinsically
and inversely proportional to the ability of the solid to deform (or
distort) the input sequence and make it less random, i.e., more random-deficient.

The current paper presents first evidence that validate the model
of \cite{Ratsaby_entropy}. We choose to simulate a solid structure
which consists of a one-dimensional vibrating solid-beam to which
we apply a random input force sequence and observe the displacement
of the beam at its other end for a finite interval of time. We determine
empirically the relationship between the algorithmic complexity of
the structure to the stochasticity of the output response. The relationship
confirms the theory of \cite{Ratsaby_entropy}.

The paper is organized as follows: in section \ref{sec:Algorithmic-Complexity}
we give a brief introduction to the main concepts of the area of algorithmic
complexity and randomness. In section \ref{sec:Aim-of-the} we state
concisely the aim of the paper. In section \ref{sec:The-solid's-equations}
we develop the equations that describe the solid deformations and
compute the numerical equations needed to produce the computer simulation
of the solid's response to external forces. In section \ref{sec:Experimental-results}
we state the experimental setup, results and analysis. In section
\ref{sec:Conclusions} we state the conclusions.

\section{\label{sec:Algorithmic-Complexity}Background}

Kolmogorov \cite{Kolmogorov65} proposed to measure the conditional
complexity of a finite object $x$ given a finite object $y$ by the
length of the shortest binary sequence $\pi$ (a program for computing
$x$) which consists of $0$s and $1$s and which reconstructs $x$
given $y$. Formally, this is defined as

\begin{equation}
K(x|y)=\min\{\ell(\pi):\phi(\pi,y)=x\}\label{eq:K}\end{equation}
where $\ell(\pi)$ is the length of the sequence $\pi$, $\phi$ is
a universal partial recursive function which acts as a description
method, i.e., when provided with input $(\pi,y)$ it gives a specification
for $x$. The word universal means that the function $\phi$ can emulate
any Turing machine (hence any partial recursive function). One can
view $\phi$ as a universal computer that can interpret any programming
language and accept any valid program $\pi$. The Kolmogorov complexity
of $x$ given $y$ as defined in (\ref{eq:K}) is the length of the
shortest program that generates $x$ on this computer given $y$ as
input. The special case of $y$ being the empty binary sequence gives
the unconditional Kolmogorov complexity $K(x)$. 

Let $\Xi$ be the space of all finite binary sequences and denote
by $\Xi_{n}$ the set of all finite binary sequences of length $n$.
An admissible \emph{selection rule} $R$ \cite{Vyugin99} is a partial
recursive function on $\Xi$ that picks certain bits from a binary
sequence $x$. Let $R(x)$ denote the selected subsequence. By $K(R|n)$
we mean the length of the shortest program computing the subsequence
$R(x)$ given $n$. Kolmogorov introduced a notion of randomness deficiency
$\delta(x|n)$ of a finite sequence $x\in\Xi_{n}$ where $\delta(x|n)=n-K(x|n)$
and $K(x|n)$ is the Kolmogorov complexity of $x$ not accounting
for its length $n$, i.e., it is a measure of complexity of the information
that codes only the specific pattern of $0$s and $1$s in $x$ without
the bits that encode the length of $x$ (which is $\log n$ bits).
Randomness deficiency measures the opposite of chaoticity of a sequence.
The more regular the sequence the less complex (chaotic) and the higher
its deficiency. An infinitely long binary sequence is regarded random
if it satisfies the principle of stability of the frequency of $1$s
for any of its subsequences that are obtained by an admissible selection
rule \cite{Kolmogorov63,Kolmogorov98}.

In \cite{Kolmogorov65} it was shown that the stochasticity of a finite
binary sequence $x$ may be precisely expressed by the deviation of
the frequency of ones from some $0<p<1$, for any subsequence of $x$
selected by an admissible selection rule $R$ of finite complexity
$K(R|n)$. The chaoticity of $x$ is the opposite of its randomness
deficiency, i.e., it is large if its Kolmogorov complexity is close
to its length $n$. The works of \cite{Kolmogorov65,Asarin87,Asarin88,Vyugin99}
relate this chaoticity to stochasticity. In \cite{Asarin87,Asarin88}
it is shown that chaoticity implies stochasticity. This can be seen
from the following relationship (with$p=\frac{1}{2}$):

\begin{equation}
\left|\nu(R(x))-\frac{1}{2}\right|\leq c\sqrt{\frac{\delta(x|n)+K(R|n)+2\log K(R|n)}{\ell(R(x))}}\label{Ineq}\end{equation}
 where for a binary sequence $s$, we denote by $\nu(s)=\frac{\#(s)}{\ell(s)}$
the frequency of $1$s in $s$ where $\#(s)$ denotes the number of
$1$s in $s$, and $\ell(R(x))$ is the length of the subsequence
selected by $R$, $c>0$ is some absolute constant. From this we see
that as the chaoticity of $x$ grows (randomness deficiency decreases)
the stochasticity of the selected subsequence grows (bias from $\frac{1}{2}$
decreases). The information content of the selection rule, namely
$K(R|n)$, has a direct effect on this relationship: the lower $K(R|n)$
the stronger the stability (smaller deviation of the frequency of
$1$s from $\frac{1}{2}$).

\section{\label{sec:Aim-of-the}Aim of the paper}

In this paper we provide first evidence that the basic notion of randomness
and its relationship to complexity (as discussed in the previous section)
underlie the behavior of physical systems. This supports the ideas
introduced in \cite{Ratsaby_entropy}. We focus on a system composed
of a vibrating elastic solid (described by the classical equations
of solid mechanics) and its interaction with a random input force.
We show that as a result of this interaction, the deformation of the
solid over time can be described as an output sequence whose stochastic
and algorithmic properties follow those of an output subsequence selected
by a selection rule of a finite complexity. Based on a large sample
of computer-generated simulations of such solids we provide statistically
significant results that show that the complexity of the system inversely
affects the complexity of the solid deformations (observed output)
and its stochasticity agrees with the theory (\ref{Ineq}). The next
section describes the solid's mechanical equations.

\section{\label{sec:The-solid's-equations}The solid's equations}

The solid consists of an elastic homogeneous and one-dimensional beam
of length $L$. Let us denote by $x$ the position on the beam so
that $0\leq x\leq L$ and by $\overrightarrow{x}$ the unit vector
on the $x$-axis. Denote by $\overrightarrow{f}=f(x,t)\overrightarrow{x}$
a force applied at time $t$ on position $x$ in the direction of
$\overrightarrow{x}$. We define by$\overrightarrow{u}=u(x,t)\overrightarrow{x}$
the displacement at time $t$ on $x$. The classical equation which
describes the field of displacements $u$ at a specific position and
time when a force $f$ is applied is as follows:

\begin{eqnarray}
\left(\frac{\partial^{2}u}{\partial t^{2}}-\frac{E}{\rho}\frac{\partial^{2}u}{\partial x^{2}}\right)(x,t) & = & f(x,t),(0<x<L,t>0),\label{eq:solid1}\end{eqnarray}
where $E$ is Young's modulus (the ratio of stress to corresponding
strain when the beam behaves elastically), and $\rho$ is the mass
density. We impose the following boundary conditions:

\begin{equation}
u(0,t)=u(L,t)=0,\:\forall t>0,\label{eq:solid2}\end{equation}
i.e., the beam is fixed at its two ends so the only displacements
is due to internal elasticity stresses of the material. Let $u_{0}(x),u_{1}(x)$
be two given functions that satisfy $u_{0}(0)=u_{0}(L)=0$. As initial
conditions we set the following,

\begin{eqnarray}
u(x,0) & = & u_{0}(x),0<x<L,\label{eq:solid3}\\
\frac{\partial u}{\partial t}(x,0) & = & u_{1}(x),0<x<L.\label{eq:solid4}\end{eqnarray}
Equations (\ref{eq:solid1}-\ref{eq:solid4}) represent the model
that describes the deformations of the elastic solid. In order to
simulate the response of the solid to external forces we use the following
numerical approximation. This is performed by introducing a regular
mesh of the $[0,L]$ interval with a constant step $\triangle x$
such that $N+2$ equally spaced points are distributed on $[0,L]$.
Specifically, we have the following mesh: $x_{0}=0$, $x_{i}=x_{i-1}+\triangle x,$
$1\leq i\leq N+1$, $x_{N+1}=L$ and $\triangle x=\frac{L}{N+1}$. 

Similarly, if time $t$ belongs to the interval $[0,T]$, we introduce
$M+1$ discrete time points $t_{0}=0$, $t_{n}=n\triangle t$, $1\leq n\leq M$,
where $\triangle t=\frac{T}{M}$. Let us introduce the approximation
sequence $\widetilde{u}(j,n)$, $1\leq j\leq N$ and $1\leq n\leq M$
such that $\widetilde{u}(j,n)\approx u(x_{j},t_{n})$, where $u$
is solution of (\ref{eq:solid1}-\ref{eq:solid4}). 

Let us also denote by $\widetilde{f}(j,n)\equiv f(x_{j},t_{n})$.
We consider the following finite differences scheme to get an approximation
of (\ref{eq:solid1}-\ref{eq:solid4}):

\begin{flushleft}
\begin{eqnarray}
\widetilde{u}(j,n+1) & = & 2\tilde{u}(j,n)-\widetilde{u}(j,n-1)+(\triangle t)^{2}\widetilde{f}(j,n)\nonumber \\
 &  & +\left(\frac{\triangle t}{\triangle x}\right)^{2}\frac{E}{\rho}\left[\widetilde{u}(j+1,n)-2\widetilde{u}(j,n)+\widetilde{u}(j-1,n)\right],\label{eq:diff1}\\
\widetilde{u}(0,n) & = & \widetilde{u}(N+1,n)=0,\label{eq:diff2}\\
\widetilde{u}(j,0) & = & u_{0}(x_{j}),\label{eq:diff3}\\
\widetilde{u}(j,1) & = & u_{0}(x_{j})+(\triangle t)u_{1}(x_{j})\label{eq:diff4}\end{eqnarray}
 $ $provided that the following CFL stability condition on the solid's
parameters is satisfied,\[
\sqrt{\frac{E}{\rho}}\frac{\triangle t}{\triangle x}\leq1.\]
 In the next section we describe the experimental setup and results
produced by (\ref{eq:diff1}-\ref{eq:diff4}). 
\par\end{flushleft}

\begin{flushleft}

\par\end{flushleft}

\section{\label{sec:Experimental-results}Experimental results}

We performed a series of experiments which consisted of several hundreds
simulation trials of the response of a vibrating solid (henceforth
called a \emph{system}) to an input force sequence. We used the numerical
equations of section \ref{sec:The-solid's-equations} as the solid's
model. As a choice of  parameters we took $L=20$, $T=70$, $E=0.7$,
$\rho=0.4$, $N=30$, $M=200$.

A system consists of a solid whose length is divided into 31 positions,
$0,1,\ldots,30$. A force sequence $\widetilde{f}(15,n)$ is applied
at position $15$ while for all remaining positions the applied force
is of zero magnitude. The non-zero force sequence $\widetilde{f}(15,n)$
makes the solid vibrate \emph{a priori} hence we call the system a
\emph{vibrating solid}. This force sequence consists of a series of
ternary values $-1,0,+1$ scaled by a constant of $30$. The length
of the sequence is $200$ and the symbols are obtained sequentially
by a repeated series of random draws using the random variable $\mathcal{F}$
with the following probability distribution: let $0<p\leq1$, then
$\mathcal{F}$ takes the value $0$ with probability $1-p$, the value
$+1$ with probability $\frac{p}{2}$, and $-1$ with probability
$\frac{p}{2}$. The complexity of the sequence is controlled by the
choice of $p$. We used a different $p$ for different trials by randomly
picking its value and using it as the parameter value $p$ of the
distribution of the random variable $\mathcal{F}$.

To the system we apply an input force sequence $\widetilde{I}(1,n)$
at position $1$ consisting of $200$ randomly drawn binary values
$+1$ and $-1$ each with probability $\nicefrac{1}{2}$ and scaled
by a constant $10$. Note that this input force is applied to a vibrating
solid (as mentioned above). As the output of the system, we observe
the displacement sequences at five positions $\widetilde{u}(N-5,n)$,
$\ldots$, $\widetilde{u}(N-1,n)$, $1\leq n\leq200$ and convert
their values $a$ from real to ternary $V(a)$ using the following
rule: given $a\in\mathbb{R}$ then $V(a)=+1$, $0$ and $-1$ if $a>\tau$,
$|a|\leq\tau$ and $a<-\tau$, respectively, with $\tau=0.1$. We
then append these five ternary sequences together to form a single
ternary output sequence of length $1000$ (henceforth this is called
the \emph{output} \emph{sequence}). We also consider the subsequence
of this output sequence which consists only of the values $+1$, $-1$,
i.e., without the zeros (we call this the \emph{output} \emph{subsequence}).

As an estimate of the complexity $K(x)$ of a sequence $x$ we use
a standard compression algorithm (Gzip, which is a variation of the
algorithm of \cite{LZ77}) to compress $x$. The length of the resulting
compressed version of $x$ is used as an approximation of $K(x)$.
Henceforth, when we say system complexity we mean the length of the
compressed version of the sequence consisting of all applied forces
appended sequentially into one string with $31\cdot200=6200$ ternary
symbols (in our experiments, all but the $\widetilde{f}(15,n)$ force
are just all-zeros hence in this $6200$-long string approximately
only $200$ bits contain information). The output complexity is the
length of the compressed version of the ternary output sequence.

\begin{figure}[h]
\begin{raggedright}
\subfigure{\includegraphics[clip]{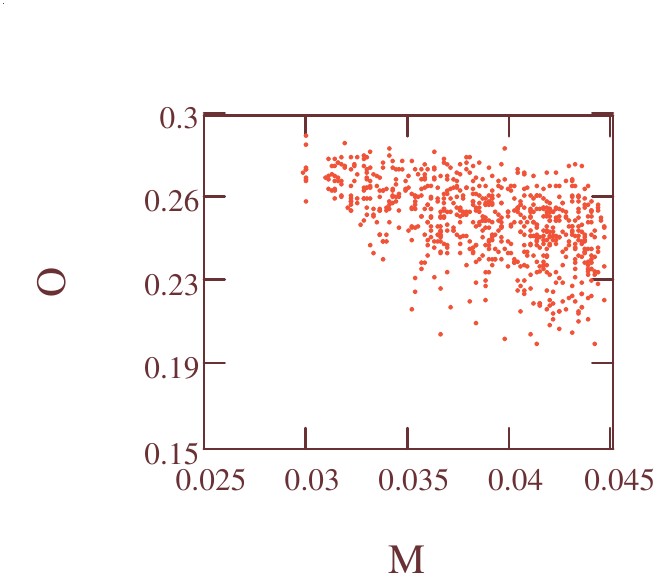}}\subfigure{\includegraphics[scale=1.09]{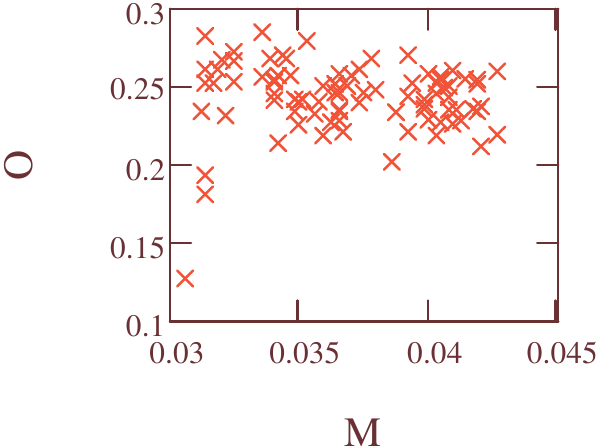}} 
\par\end{raggedright}

\caption{\label{fig:Output's-comp vs M}Output's complexity O versus the  complexity
M, (a) with random force input, (b) with no input}

\end{figure}

Let $M$ denote the ratio of the compressed length divided by the
uncompressed length of the system and let $O$ denote this ratio for
the output sequence. A large $M$ (or $O$) means that the compressed
length is larger hence the complexity of the system (or output sequence)
is larger. We sometime simply refer to $M$ and $O$ as the system
and output complexity, respectively. Figure \ref{fig:Output's-comp vs M}
displays two sets of trials.

In each trial of set (a) a random input force sequence was applied
at position $1$ (as described above). In each trial of set (b) \emph{no}
input sequence was applied. As is seen, the resulting behavior is
clearly different in each of the two sets of trials. With an input
present, as the  complexity $M$ increases there appears to be a decreasing
trend in the value of $O$ and an increase in the spread, i.e., the
range of possible values of $O$. With no input, both $O$ and its
spread of values are basically constant with respect to $M$. 

\begin{figure}[h]
\begin{centering}
\includegraphics{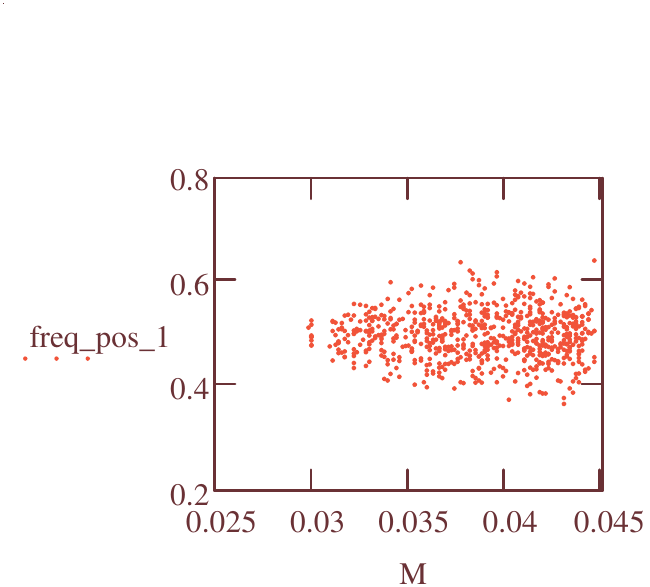} 
\par\end{centering}

\caption{\label{fig:Output-frequeny-of}Output frequeny of $1$s versus $M$}

\end{figure}

In Figure \ref{fig:Output-frequeny-of} we plot the frequency of $1$s
in the output subsequence (this is the number of $1$s divided by
the number of non-zero symbols in the output sequence).

As can be seen, with an increase in the  complexity there appears
to be an increase in the spread of possible frequency values. Before
we further discuss these results we proceed to perform the statistical
tests.

\subsection{Analysis}

\begin{figure}[h]
\includegraphics[scale=0.6]{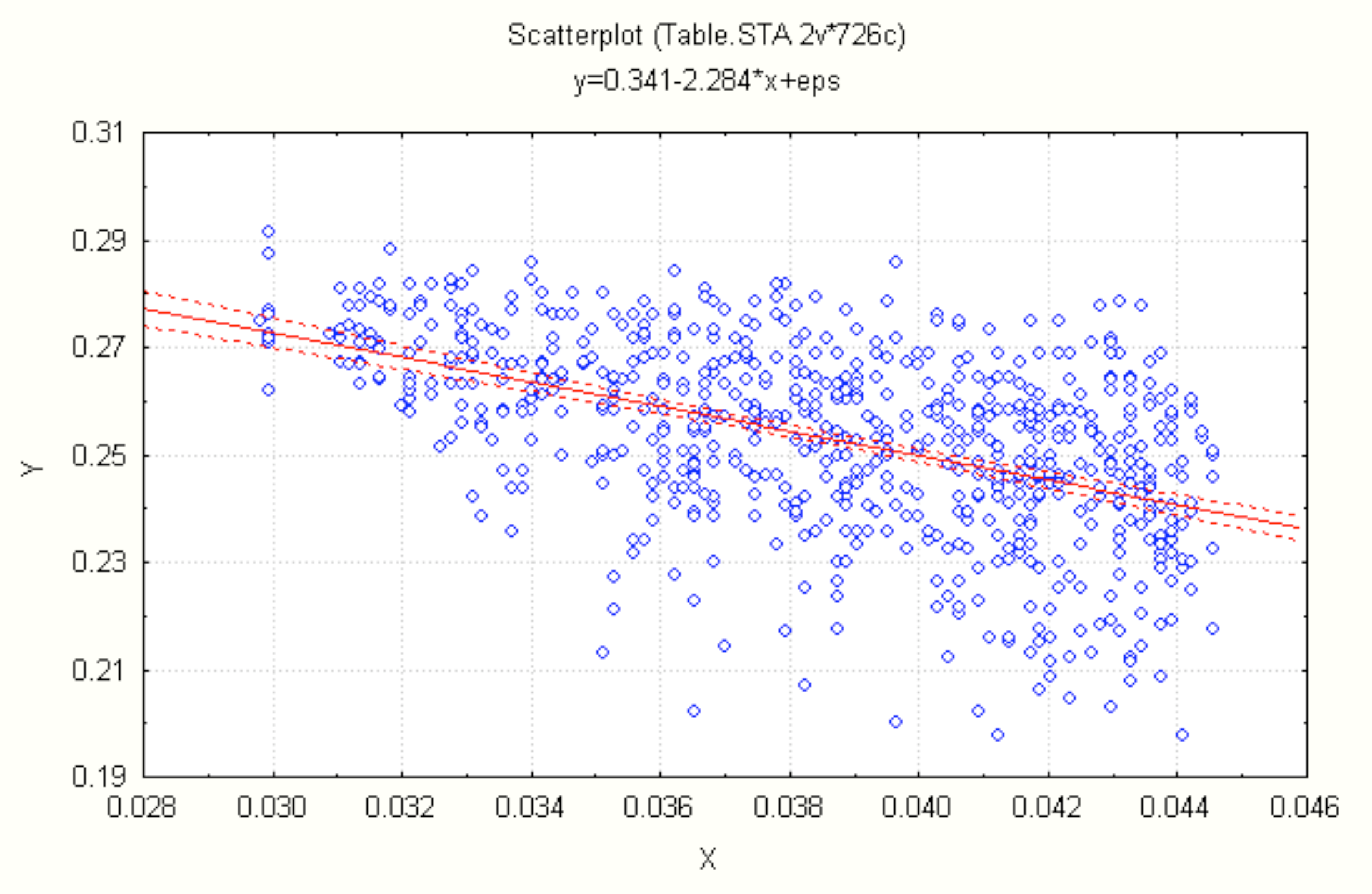} 

\caption{\label{fig:Estimate-for-Y}Estimate for output complexity $Y$ as
a function of  complexity $X$}

\end{figure}

In order to test the significance of these results we estimate the
output complexity O as a function of the  complexity $M$. Denote
by $X$ and $Y$ the random variables corresponding to $M$ and $O$,
respectively. Let the underlying conditional probability distribution
function be $P(Y|X)$ with marginals $P(X)$, $P(Y)$. As a sample
we use the set of trials of Figure 1(a), denoted by $S=\left\{ \left(x_{i},y_{i}\right)\right\} _{i=1}^{N}$
with cardinality $N=723$, and do linear regression in order to estimate
$Y$ with dependence on $X$. Figure \ref{fig:Estimate-for-Y} shows
the resulting estimate, \begin{equation}
\hat{Y}(X)=0.341-2.284\, X,\label{eq:yhat}\end{equation}
 surrounded by the $95\%$ confidence limits for the regression line,
i.e., the actual regression line of the population falls within the
limits defined by the two curved dashed lines.

\begin{figure}[h]
\includegraphics[scale=0.6]{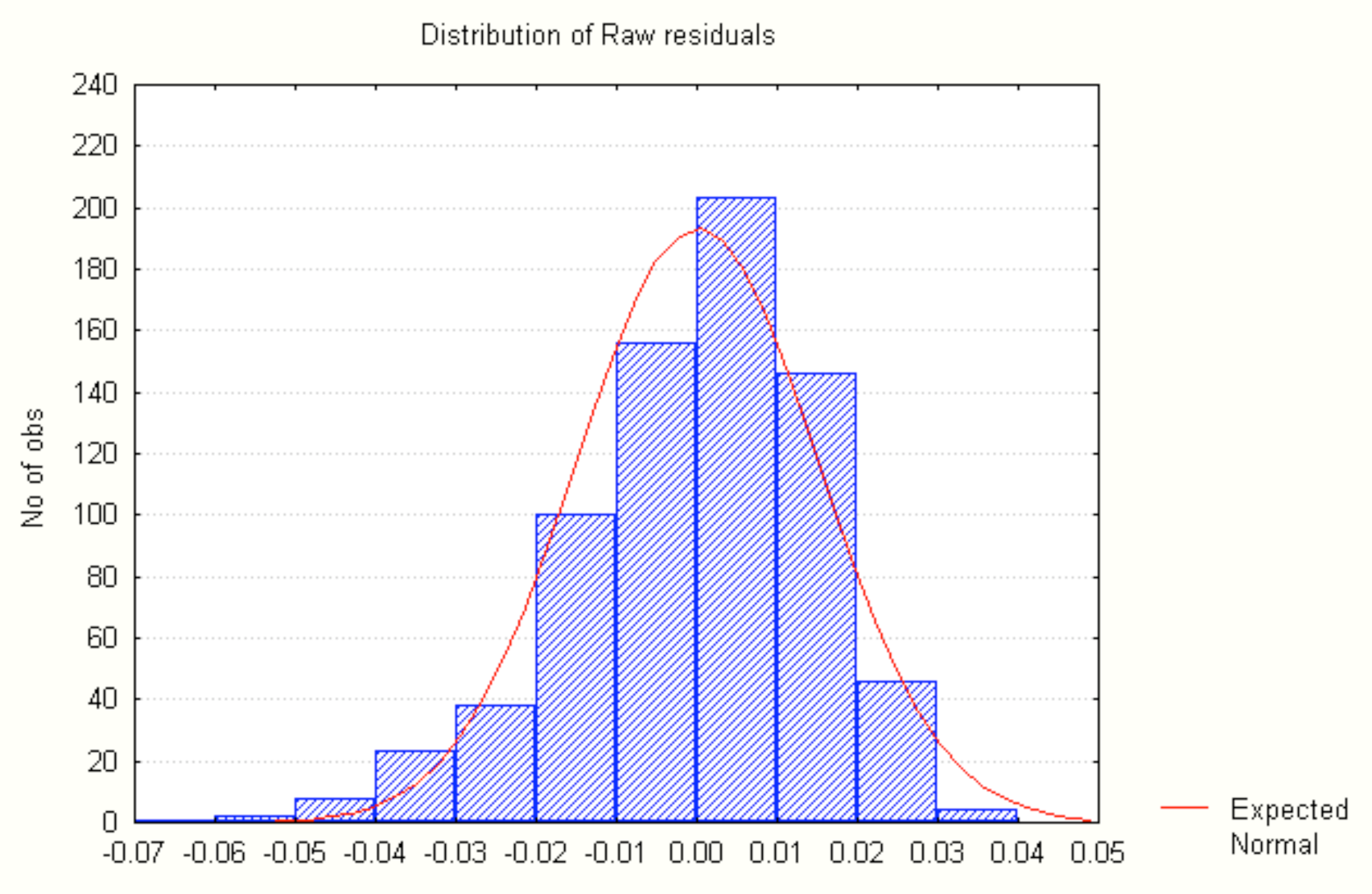} 

\caption{\label{fig:Distribution-of-the}Distribution of the residuals $y_{i}-\hat{Y}(x_{i})$}

\end{figure}

The following summarizes the accuracy of this linear regression estimate:
$R^{2}=.246187335$ is the coeeficient of determination which measures
the reduction in total variation of $Y$ due to $X$ and is defined
as $R^{2}=1-(SS_{R}/SS)$ with $SS_{R}$ $=\sum_{i}\left(y_{i}-\hat{Y}(x_{i})\right)^{2}$
being the sum of squares of the residuals, $SS=\sum_{i}\left(y_{i}-\bar{y}\right)^{2}$
the total variation and $\bar{y}=\frac{1}{N}\sum_{i}y_{i}$ . The
square root $R$ is the coeeficient of correlation between the independent
variable $X$ and dependent variable $Y$. The standard error $SE=.015128505$
where $SE=\sqrt{\frac{1}{N}SS_{R}}$. Dividing $SS_{R}$ and $SS$
by their degrees of freedom and taking their ratio $F=SS/SS_{R}$
as an overall $F$ test gives $F(1,724)=236.4508$ which amounts to
a $p$-value less than $0.000000$. Thus with very high confidence
the residual variance differs from the total variation hence the linear
estimate $\hat{Y}(X)$ explains well the variation of $Y$. The distribution
of the residuals (shown in Figure \ref{fig:Distribution-of-the})
is very close to the normal distribution and the Durbin-Watson $d$
value is $1.994$ which implies that the assumptions on the residuals
being uncorrelated and normaly distributed are met.

\begin{figure}[h]
\includegraphics[scale=0.6]{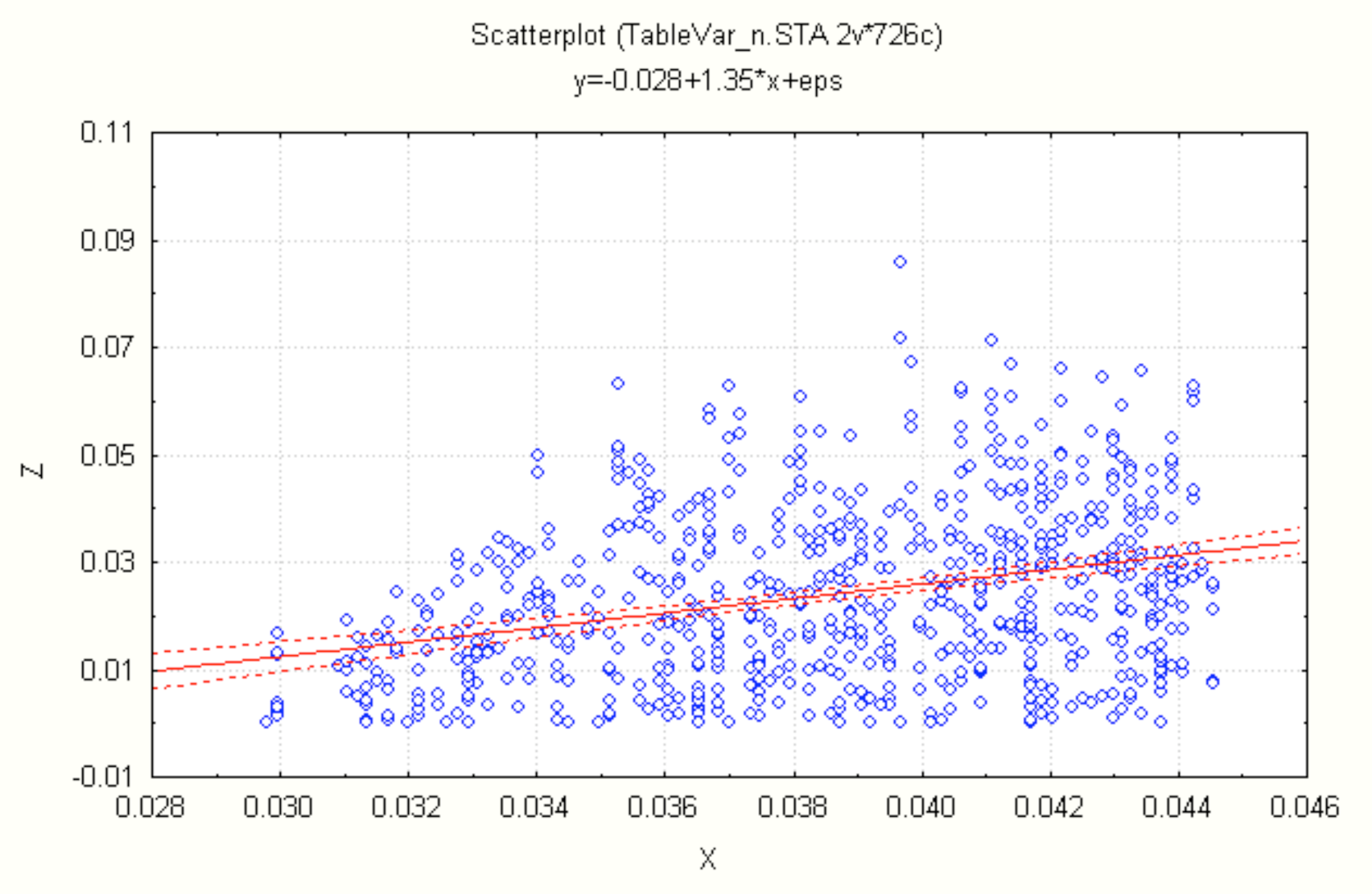} 

\caption{\label{fig:Estimate-of-the spread}Estimate of the spread in output
complexity $Z$ as a function of the system's complexity $X$}

\end{figure}

\begin{figure}[h]
\includegraphics[scale=0.6]{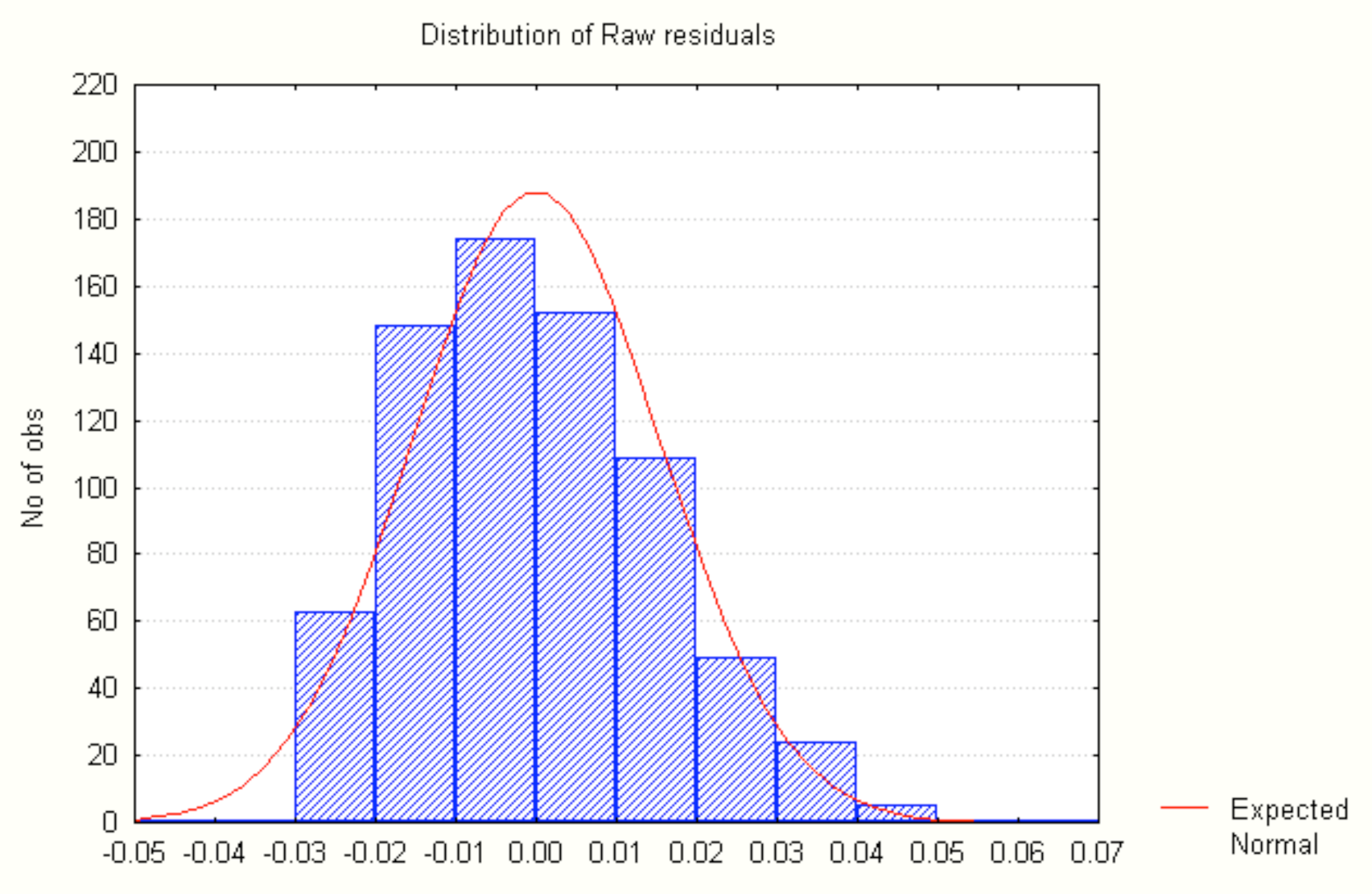} 

\caption{\label{fig:Distribution-of-Zresiduals}Distribution of residuals for
$z_{i}-\hat{Z}(x_{i},y_{i})$}

\end{figure}

Next, from Figure 1(a) it is evident that as the  complexity $X$
increases the spread of the output complexity $Y$ increases. To quantify
this assertion let us represent this spread by the random variable
\begin{equation}
Z(X,Y)=Y-\min_{y:P(y|X)>0}y.\label{eq:z(xy)}\end{equation}
As we have done above for $Y$ we now estimate $Z$ with dependence
on $X$ (the model is shown only the value of $X$ and asked to predict
$Z$). We form the following sample (based on $S$), \begin{equation}
\zeta=\left\{ \left(x_{i},z_{i}\right)\right\} _{i=1}^{N},\: z_{i}=y_{i}-\min_{x_{j}\in NN(x_{i},k)}y_{j}\label{eq:zeta_sample}\end{equation}
where $NN(x,k)$ denotes the set of $k$ nearest sample point $x_{j}$
to $x$ satisfying $x_{j}\leq x$. Figure \ref{fig:Estimate-of-the spread}
shows the resulting estimate equation (based on $k=7)$, \[
\hat{Z}(X)=-0.028+1.35\, X\]
for the regression line. This verfies the increase in the value of
$Z$ (i.e., in the spread of the output complexity $Y$) as the  complexity
$X$ increases. The following summarizes the accuracy of this regression
estimate: $R^{2}=.098$, the $F$-ratio is $F(1,724)=79.548$ with
a $p$-level smaller than $.000000$. The standard error of the estimate
is $SE=.015413$ with a Durbin-Watson $d=1.938.$ Thus the estimator
$\hat{Z}(X)$ accurately captures the variability of $Z$, i.e., the
spread of output complexity $Y$. Figure \ref{fig:Distribution-of-Zresiduals}
displays the distribution of the residuals.

\begin{figure}[h]
\includegraphics[scale=0.6]{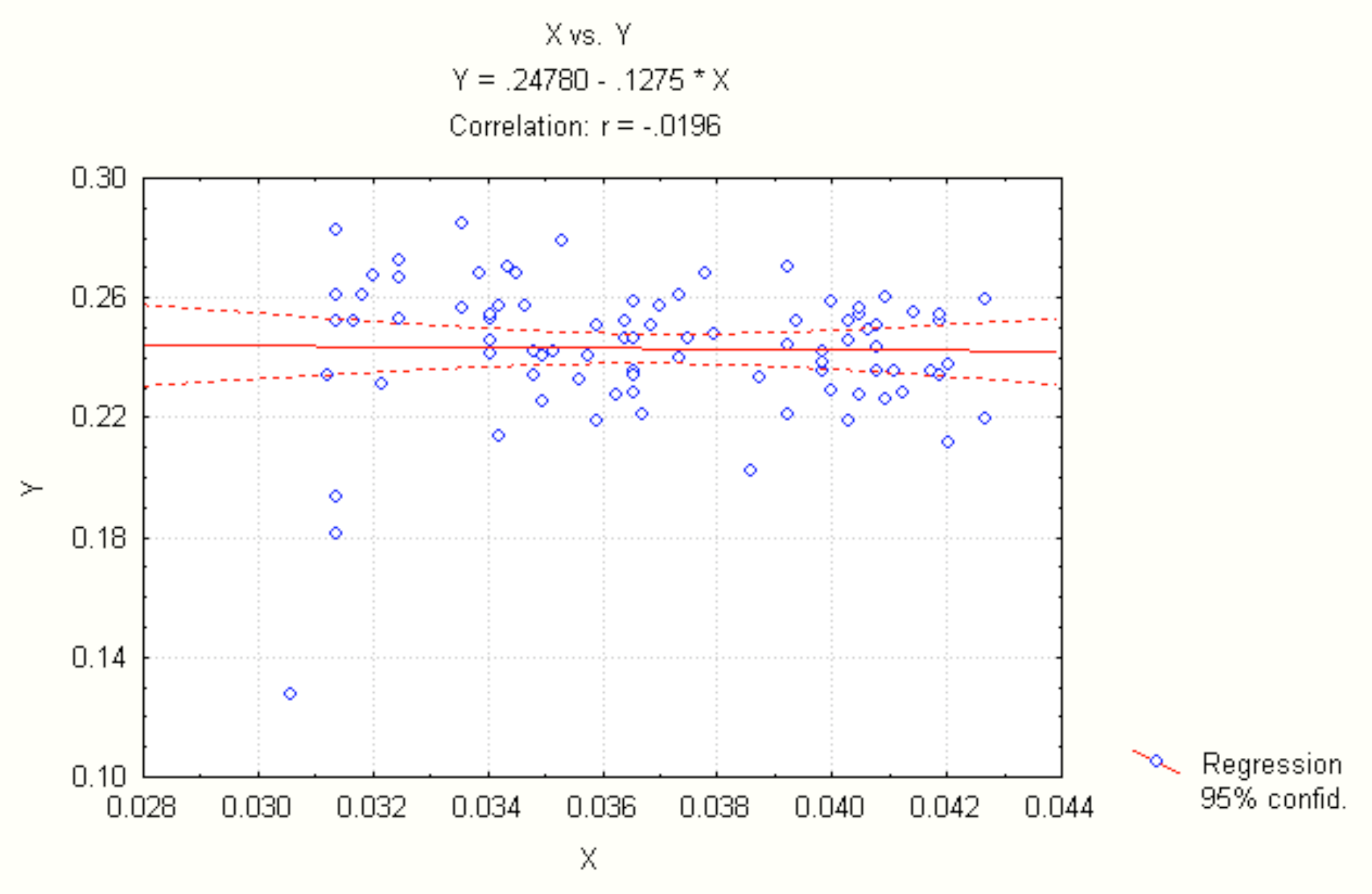} 

\caption{\label{fig:The-no-input-scatter}The no-input scatter plot of output
complexity $Y$ versus system's complexity $X$ }

\end{figure}

As mentioned above, Figure \ref{fig:Output's-comp vs M}(b) shows
that when no input is present the behavior of the output complexity
is almost unaffected by the system's complexity. To test this, we
take the set of trials used in Figure \ref{fig:Output's-comp vs M}(b)
and study the correlation between the output complexity $Y$ and the
system complexity $X$. As shown in Figure \ref{fig:The-no-input-scatter}
there is hardly any correlation between them and the slope of the
regression is almost zero.

\begin{figure}[h]
\includegraphics[scale=0.7]{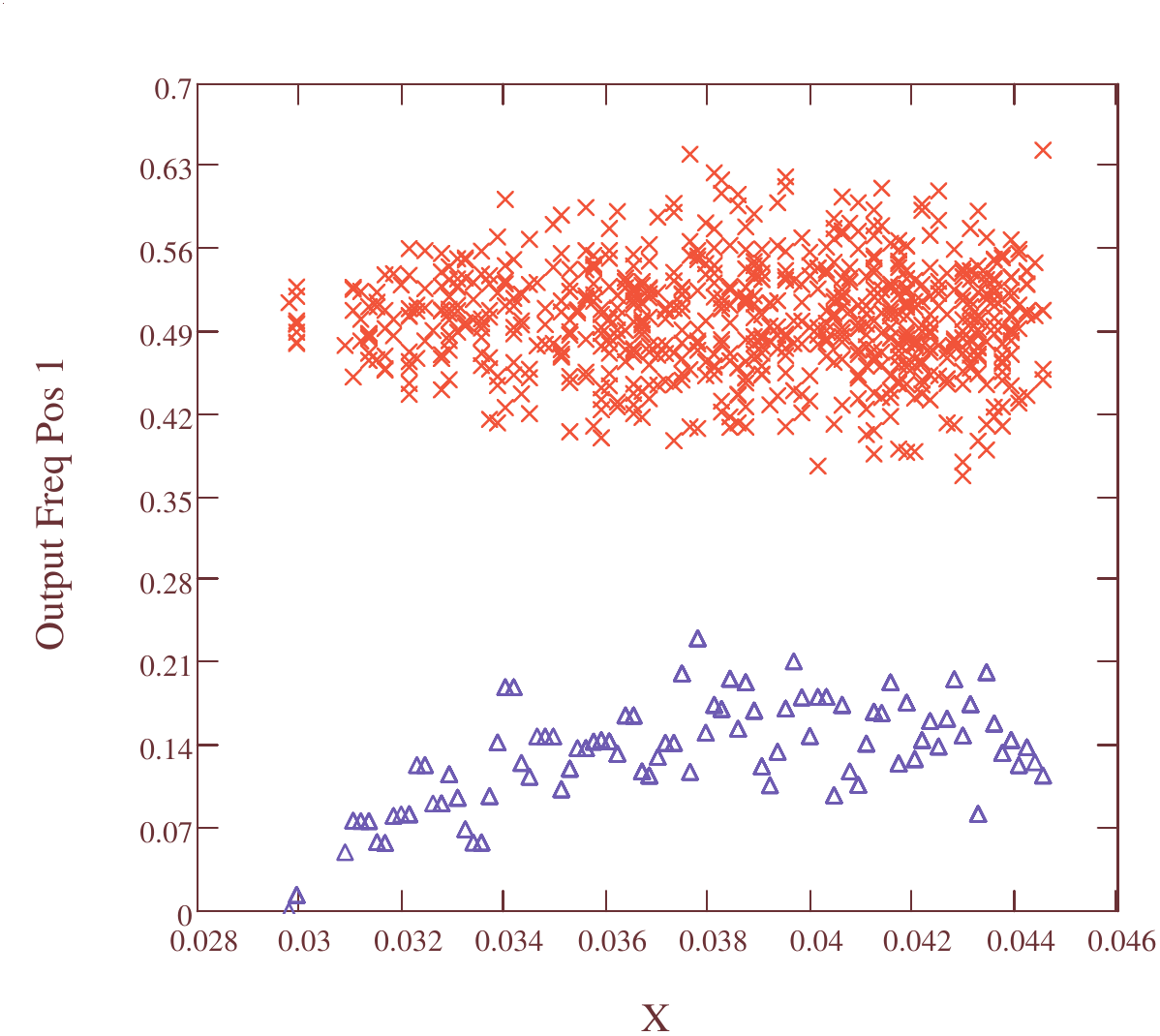} 

\caption{\label{fig:Scatter-plot-of-freq-spread}Scatter plot of frequency
of $1$s in the output subsequence (top cluster of red $\mathsf{x}s$).
The corresponding sample $\zeta'$ used to estimate the spread $W$
as a function of the system's complexity $X$ (bottom plot of blue
$\triangle$)}

\end{figure}

We already commented on the increasing spread of possible frequency
values of the output subsequence (Figure \ref{fig:Output-frequeny-of})
as the system's complexity increases (with a random input sequence
being applied). Denote by $Y$ the probability of having a $+1$ appear
in the output subsequence and let $X$ be the system's complexity.
Let us define the following random variable\[
W(X)=\max_{y:P(y|X)>0}y-\min_{y:P(y|X)>0}y\]
to represent the spread in the possible values of the probability
of $+1$. We form the following sample (based on $S$), \begin{equation}
\zeta'=\left\{ \left(x_{i},w_{i}\right)\right\} _{i=1}^{N},\: w_{i}=\max_{x_{j}\in NN(x_{i},k)}y_{j}-\min_{x_{j}\in NN(x_{i},k)}y_{j}\label{eq:sample2}\end{equation}
with $k=7.$ Figure \ref{fig:Scatter-plot-of-freq-spread} shows (on
the top scatter plot with red $\mathsf{x}$ symbols) the frequency
of $1$s in the output subsequence versus the system complexity $X$.
The bottom plot (with blue $\triangle$) shows the sample $\zeta'$
with the $w$ component on the same vertical axis.

\begin{figure}[h]
\includegraphics[scale=0.6]{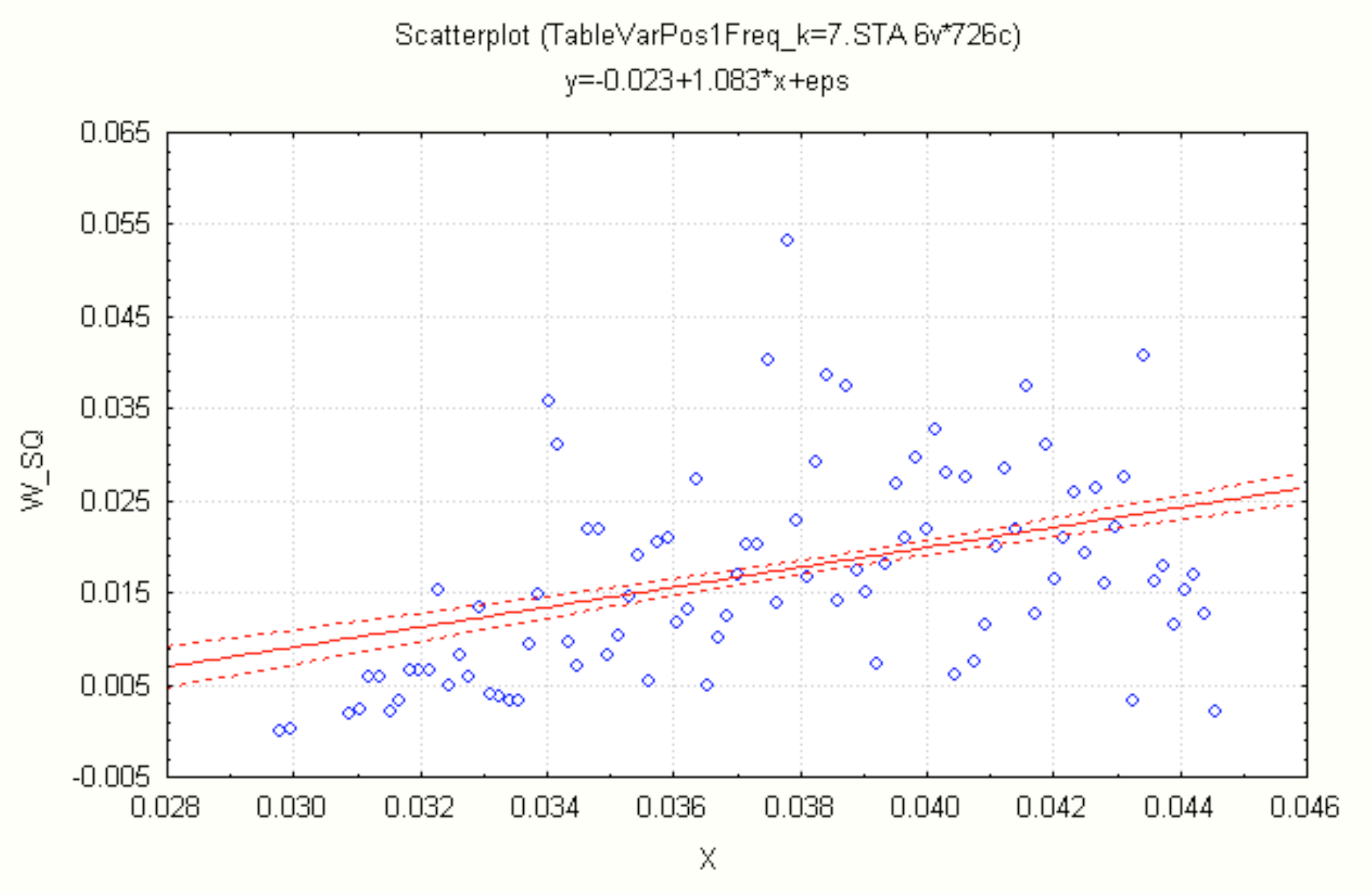} 

\caption{\label{fig:Estimate-of-the-Spread-Freq1s}Estimate of the square of
the spread $W^{2}$ of the output frequency of $1$s as a function
of the system's complexity $X$}

\end{figure}

\begin{figure}[h]
\includegraphics[scale=0.6]{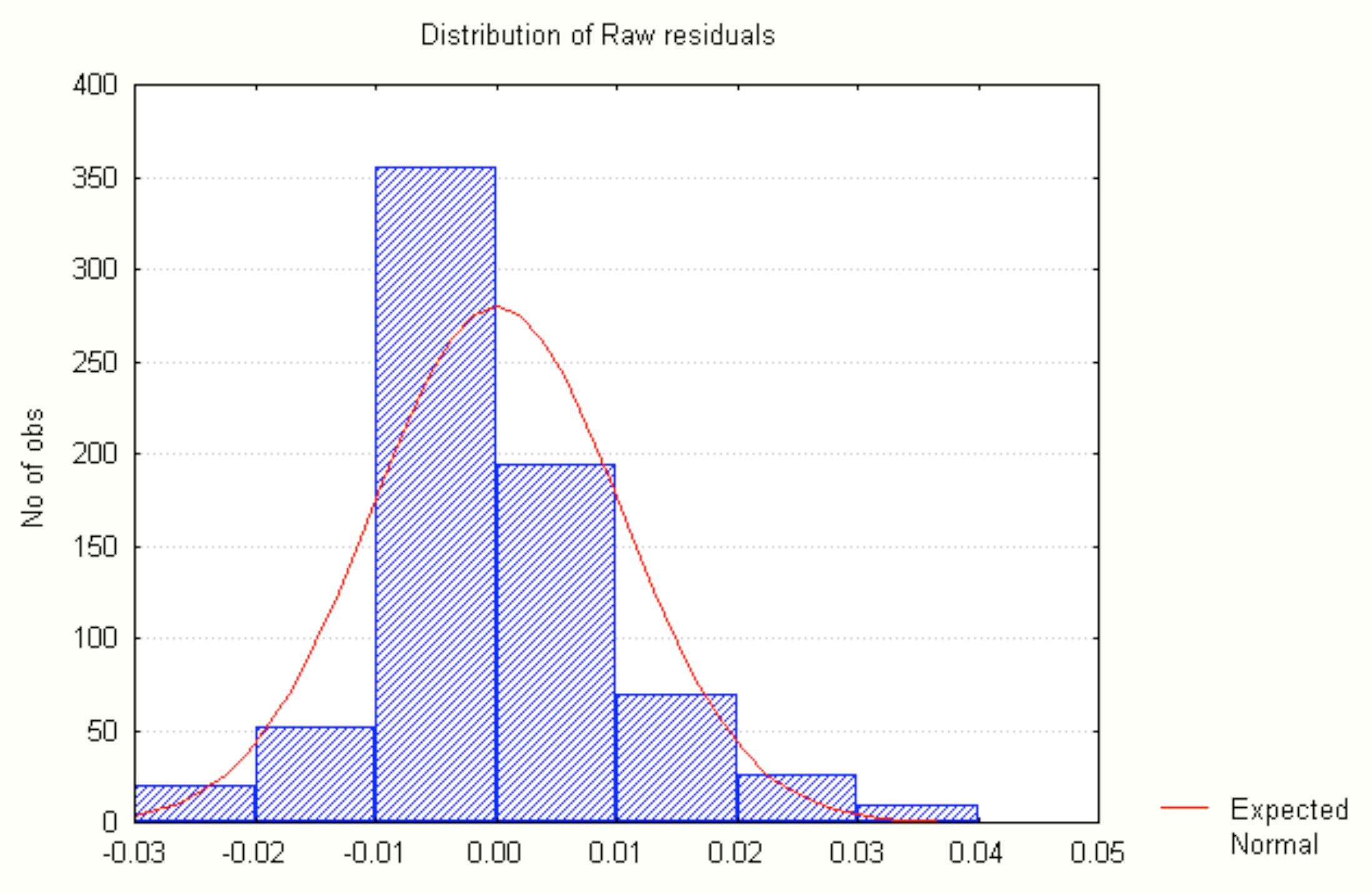} 

\caption{\label{fig:Distribution-of-residuals w}Distribution of residuals
$w_{i}^{2}-\hat{W^{2}}(x_{i})$}

\end{figure}

We estimate $W$ based on $\zeta'$ first transforming the $w_{i}$
values to $w_{i}^{2}$ and then doing linear regression to estimate
$W^{2}$. Figure \ref{fig:Estimate-of-the-Spread-Freq1s} shows the
resulting estimate equation $-0.023+1.083\, X$ for $W^{2}$. %
 It follows that the estimate of $W$ is \begin{equation}
\hat{W}(X)=\sqrt{1.083\, X-0.023}.\label{eq:what}\end{equation}
This verfies the increasing trend in the spread of values of the frequency
of $1$s as the system's complexity $X$ increases. The following
summarizes the accuracy of this linear regression estimate: $R^{2}=.135$,
the $F$-ratio is $F(1,724)=113.06$ with a $p$-level smaller than
$.00000$. The standard error of the estimate is $SE=.01037$ and
the Durbin-Watson $d=1.71$. The distribution of the residuals is
shown in Figure \ref{fig:Distribution-of-residuals w}.

\subsection{Some more details on the simulations}

Several additional graphs showing additional details of the above
experiments are shown below. Figure \ref{fig:Entropy-H-(red)} shows
the observed system description rate $M$ (scatter plot in blue) and
the entropy (the minimal expected number of bits per character) used
for the system description (red solid curve). They are plotted versus
the probability parameter $p$ (in the range $0<p\leq1)$ used to
generate the random force sequence at position $15$ of the solid.
It follows from the procedure (described above) of generating the
system's vibrating force that the entropy of the random variable $\mathcal{F}$
is $H(p)=-(1-p)\log(1-p)-p\log\frac{p}{2}.$ As seen from Figure \ref{fig:Entropy-H-(red)},
in order to get a higher system complexity one needs to draw a force
sequence with a parameter $p$ closer to $\nicefrac{1}{2}$. There
is some additional textual information ($145$ bytes) appended into
each of the files that contain the $6200$-long ternary string that
describes the system. Since $M$ is the ratio of the compressed to
uncompressed versions (the actual uncompressed length as reported
by the operating system is $6377$ bytes) then to get the rate for
the number of bits per character used to describe the system (considering
just the $145$-byte textual information and the $200$-byte force
sequence at position $15$) we multiply $M$ by $6377\cdot8$ and
divide by $345$. 

\begin{figure}[h]
\begin{centering}
\includegraphics[clip]{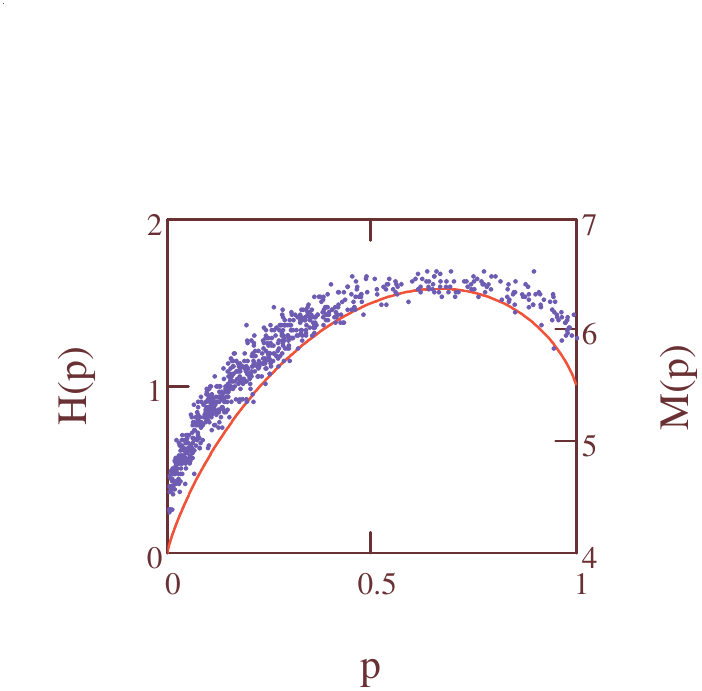} 
\par\end{centering}

\caption{\label{fig:Entropy-H-(red)}Entropy $H$ (red) v.s. observed system
complexity $M$ (blue) as a function of parameter $p$ which represents
the probability of having a non-zero (i.e. $\pm1)$ symbol in the
system's vibrating force sequence (applied at position $15$)}

\end{figure}

\begin{figure}[h]
\begin{centering}
\includegraphics[scale=0.7]{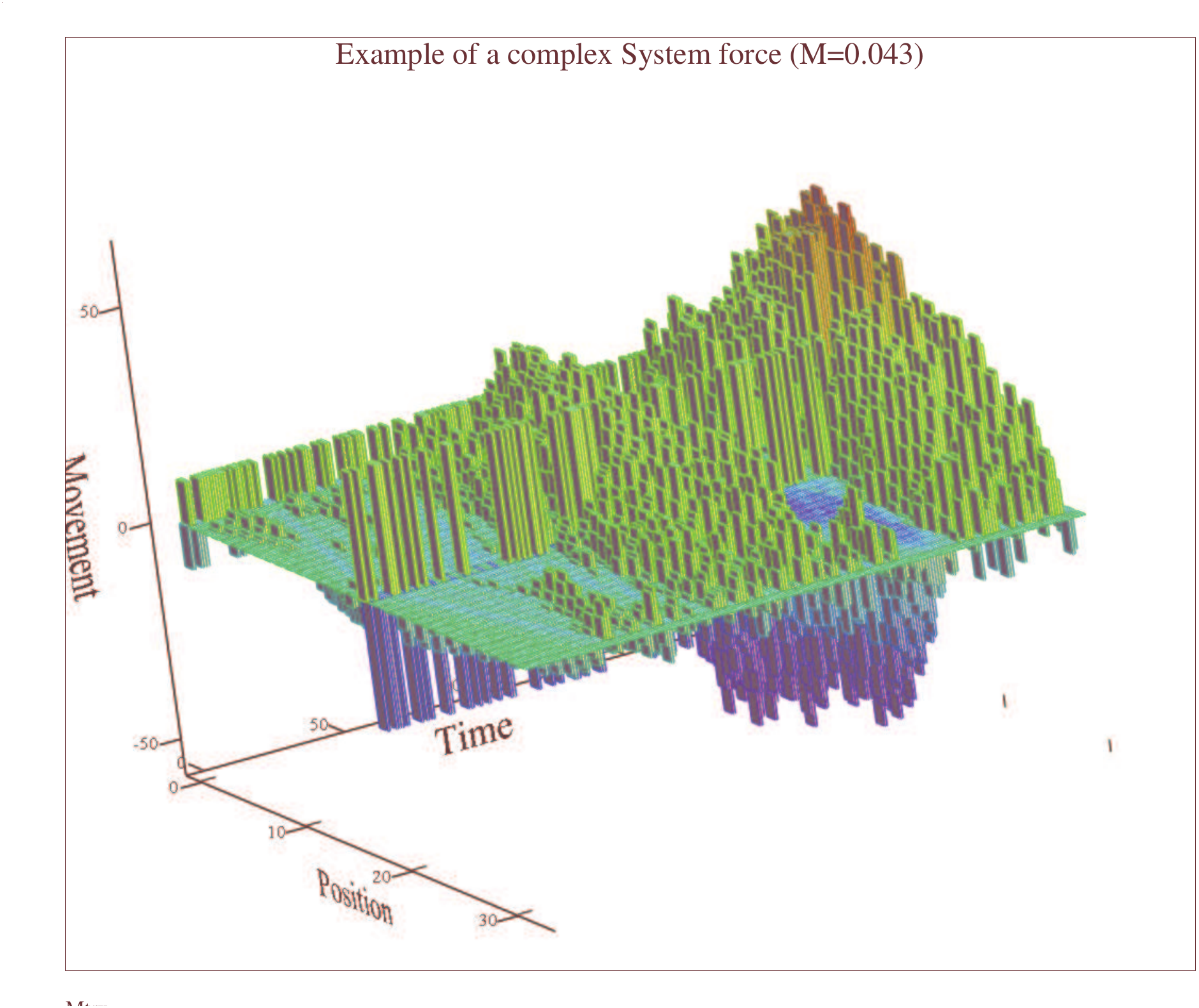} 
\par\end{centering}

\caption{\label{fig:Response-to-a1}Response to a random input force sequence
(applied at position 0). System's vibrating force (at position 15)
is of high complexity. Output is seen to be of low complexity.}

\end{figure}

\begin{figure}[h]
\begin{centering}
\includegraphics[scale=0.7]{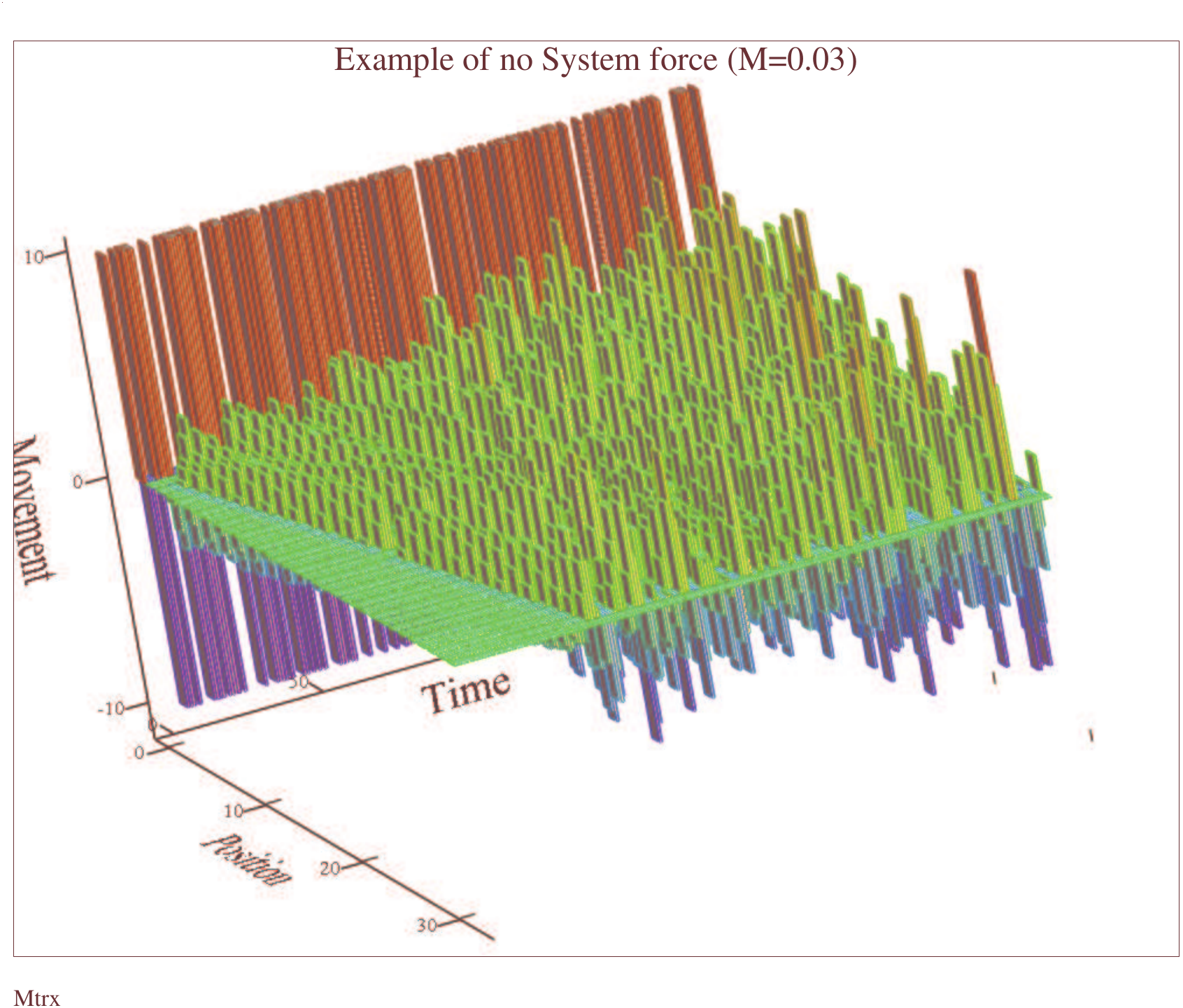} 
\par\end{centering}

\caption{\label{fig4}Response to a random input force sequence (applied at
position 0). System has \emph{no} vibrating force applied hence is
of minimal complexity. The output is seen to be of high complexity.}

\end{figure}

\begin{figure}[h]
\begin{centering}
\includegraphics[scale=0.7]{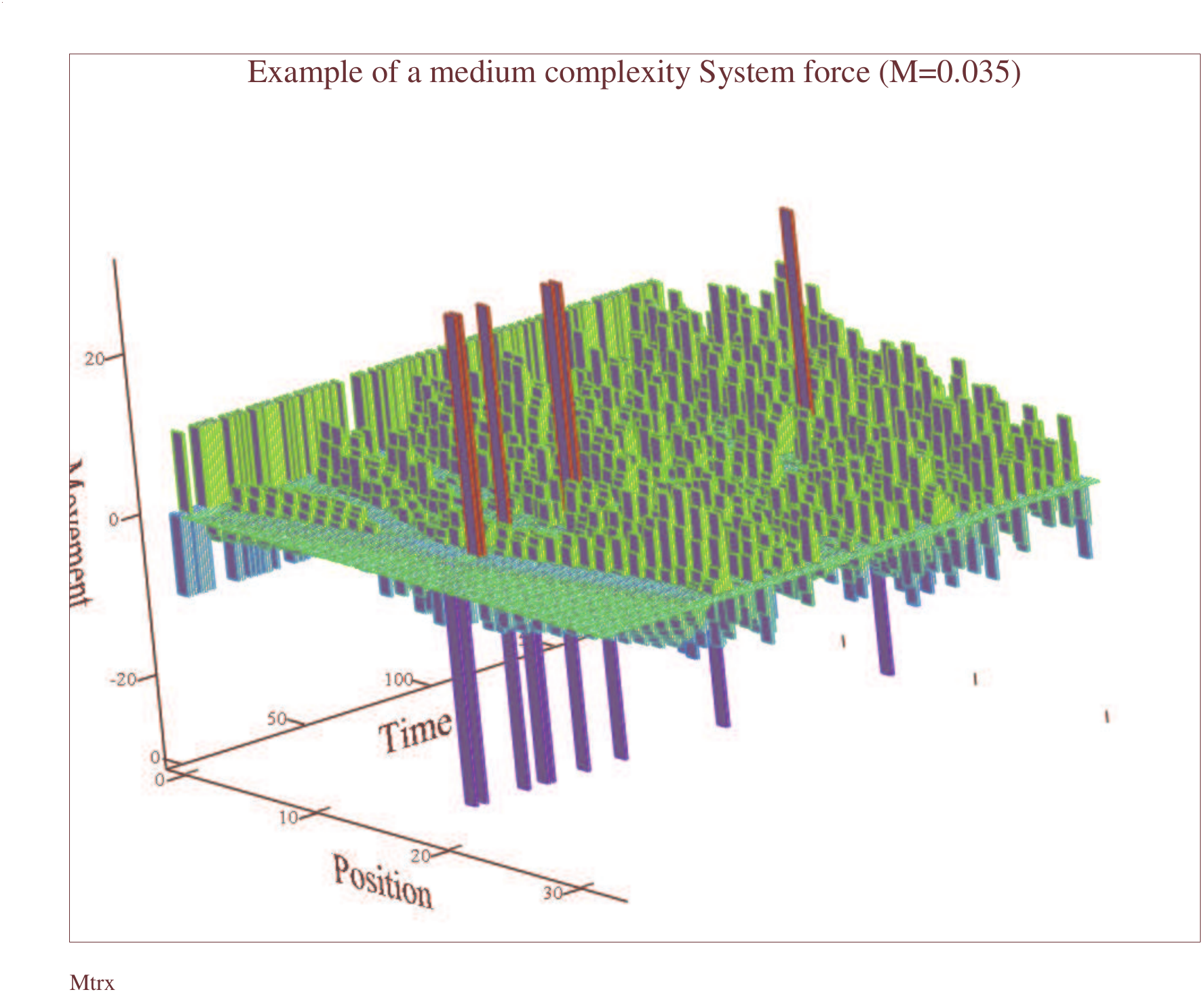} 
\par\end{centering}

\caption{\label{fig:Response-to-a2}Response to a random input force sequence
(applied at position 0). System's vibrating force (at position 15)
has a mid-level complexity. Output is seen to be of lower complexity
than the example in Figure \ref{fig4}.}

\end{figure}

\begin{figure}[h]
\begin{centering}
\includegraphics[scale=0.7]{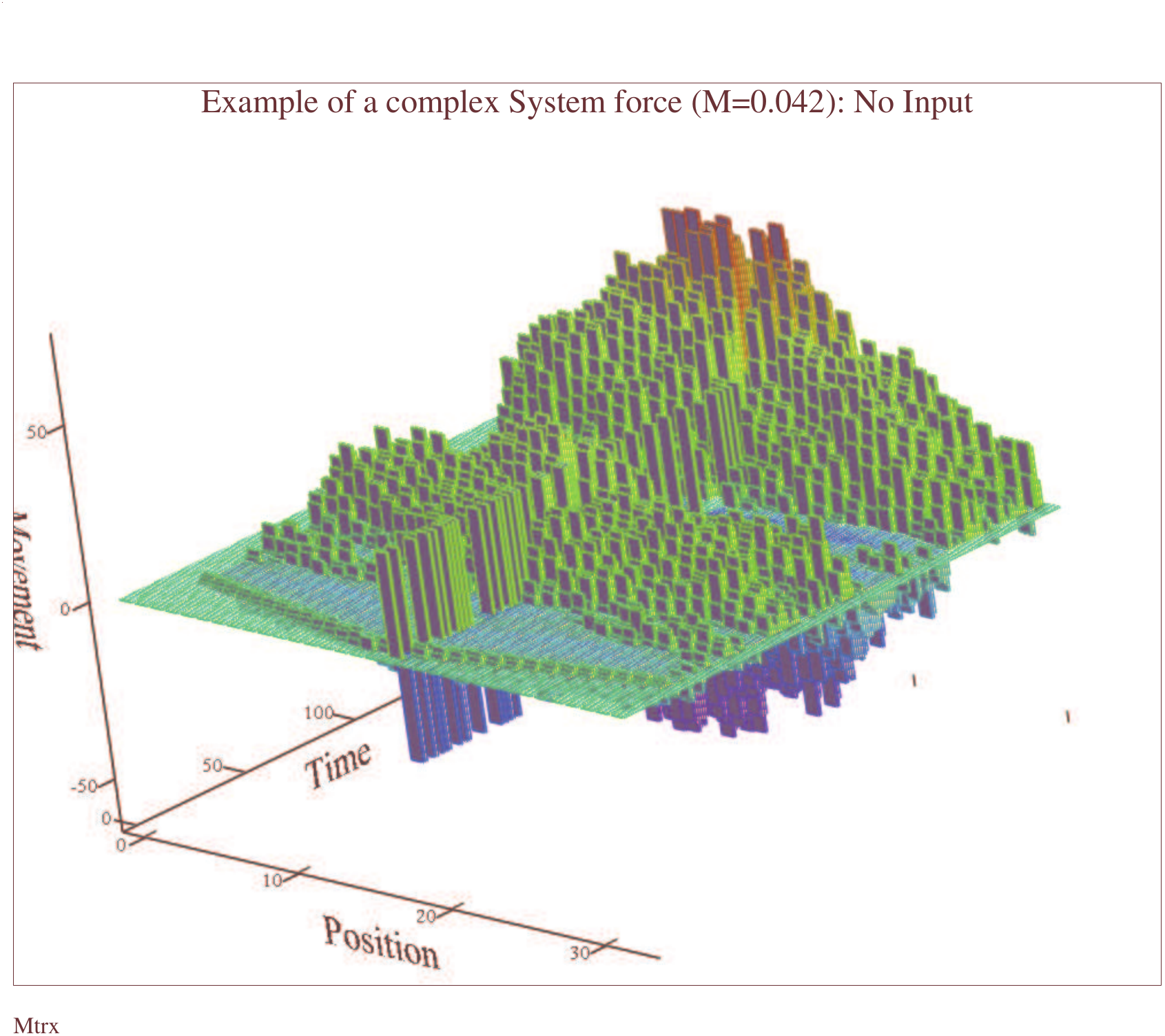} 
\par\end{centering}

\caption{\label{fig:The-no-input-response3}The no-input response to a system's
force (at position 15) of high complexity.}

\end{figure}

\begin{figure}[h]
\begin{centering}
\includegraphics[scale=0.7]{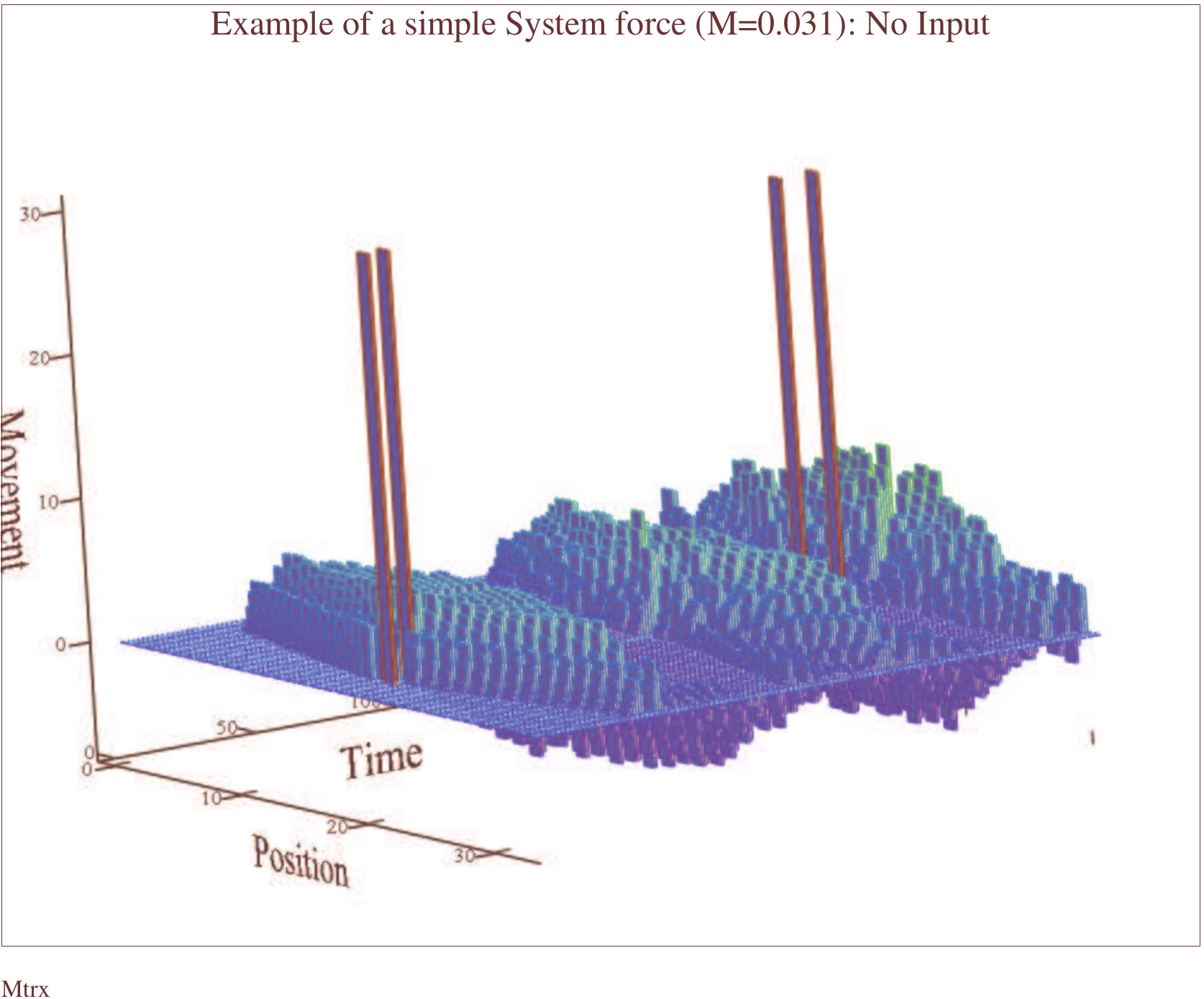} 
\par\end{centering}

\caption{\label{fig:The-no-input-response4}The no-input response to a system's
force (at position 15) of low complexity.}

\end{figure}

The next series of figures show examples of the actual solid's response
(displacement $u$ is shown on the $z$-axis) over time ($y$-axis)
along the positions of the solid ($x$-axis). In all, the input sequence
is maximally random with probability $\nicefrac{1}{2}$ for $+1$,
$-1$. The magnitude of the input force sequence is $10$ and the
magnitude of the system's vibrating force sequence is $30$. The two
force sequences are superimposed on the same 3D-plot that displays
the displacement response. The output sequence is taken as the concatenation
of the string obtained from the displacement at the last five positions
(appearing on the plot to be closest to the reader). Figure \ref{fig:Response-to-a1}
shows the response to a system of high complexity. Figure \ref{fig4}
shows a trial without a system force. This represents a low-complexity
system. Figure \ref{fig:Response-to-a2} is the response of a system
of a mid-level complexity. Figures \ref{fig:The-no-input-response3}
and \ref{fig:The-no-input-response4} show the response when no input
force is applied.

\bibliographystyle{plain} 

\section{\label{sec:Conclusions}Conclusions}

Based on these results, it is clear that when a random input sequence
is applied to the vibrating solid (system) the observed output sequence
is \emph{not} simply a result of the random vibrating force sequence
which is part of the system (applied at position 15) but is a direct
consequence of the interaction of the system with an external random
input force--when no input is present no significant correlation exists
between the system and output complexities. The strong negative correlation
between these two complexities (\ref{eq:yhat}) suggests that the
system distorts the input randomness and produces a less complex output
sequence. This agrees with the model introduced in \cite{Ratsaby_entropy}
which says that a solid effectively acts as a selection rule picking
bits from the input sequence to produce a less random output. This
is evident in the significant decrease in the output complexity (Figure
\ref{fig:Estimate-for-Y}) and increase in its spread of values (Figure
\ref{fig:Estimate-of-the spread} ) indicating that the possibility
of producing a less-complex output sequence increases as the system's
complexity rises.

The selected subsequence consists of $\pm1$ with zeroes deleted.
Being less chaotic, its stochastic level decreases. This is evident
in the increase in the square of the spread of values of the frequency
of $1$'s (Figure \ref{fig:Estimate-of-the-Spread-Freq1s}). The higher
the system complexity, the higher the spread, i.e., the larger the
bias from $\nicefrac{1}{2}$, the more the chance that the output
sequence be less chaotic and random. If we divide the compressed length
of the system by the length of the output (binary) subsequence and
denote it by $X'$ then re-estimate $W^{2}$ based on $X'$ we obtain
the following estimate for $W$,

\begin{equation}
\hat{W}(X')=\sqrt{-0.03+0.173X'}.\label{eq:xpr}\end{equation}
The $R^{2}=.0924$, $SE=.0106$, $F(1,724)=73.725$ and the p-level
less than $0.00000$. The Durbin-Watson $d=1.66$. This estimate of
the spread agrees with the rate predicted by the theory (\ref{Ineq}).
To see this, let $x$ be the input sequence, let the system be the selection rule
$R$ with a system complexity $K(R|n)$, let $R(x)$ be the output
subsequence (consisting only of binary values $\pm1$), let $\nu(R(x))$
be the frequency of $1$s in this subsequence and take the deficiency
of randomness $\delta(x|n)$ of the input sequence to be zero (since
the input sequence is maximally random). Then the theoretical rate
of the maximal possible deviation (spread) between $\nu(R(x))$ and
$\nicefrac{1}{2}$ is $O(\sqrt{K(R|n)/\ell(R(x))})$. This is the
same rate in which the estimate of spread $W$ grows with respect
to the $X'$ in (\ref{eq:xpr}).

To summarize, the results above imply that a system based on classical
equations of mechanics that consists of a vibrating solid subjected
to external random input-force acts like an algorithmic selection
rule of a finite complexity. It produces an output sequence whose
stochastic and chaotic properties are effected by the system's complexity
as predicted by the theory of algorithmic randomness. The results
confirm the model of \cite{Ratsaby_entropy}.

\end{document}